\documentclass[nopacs,twocolumn,preprintnumbers,notitlepage,amsmath,amssymb,superscriptaddress]{revtex4-2}

\usepackage[utf8]{inputenc}
\usepackage[utf8]{inputenc}
\usepackage{physics}
\usepackage{blindtext}
\usepackage{ dsfont }
\usepackage{tikz}
\usepackage{subfiles}
\usepackage{subcaption}
\usepackage{graphicx}% Include figure files
\usepackage{dcolumn}% Align table columns on decimal point
\usepackage{bm}
\usepackage{amsmath}
\usepackage{amssymb}

\newcommand{\Sim}[1]{\mathrel{\mathop{\kern 0pt\sim}\limits_{#1}}}
\newcommand{\To}[1]{\mathrel{\mathop{\kern 0pt\to}\limits_{#1}}}
\makeatletter
\AtBeginDocument{\let\LS@rot\@undefined}
\makeatother
\usepackage[]{pdfpages}

\begin{document}

\title{Aging Dynamics of $d-$dimensional Locally Activated  Random Walks }
\author{Julien Br\'emont}
\affiliation{Laboratoire de Physique Th\'eorique de la Mati\`ere Condens\'ee, CNRS/Sorbonne Université, 
 4 Place Jussieu, 75005 Paris, France}
\affiliation{Laboratoire Jean Perrin, CNRS/Sorbonne Université, 
 4 Place Jussieu, 75005 Paris, France}
\author{Theresa Jakuszeit}
\affiliation{Institut Curie, CNRS UMR 144, Universit\'e PSL, 26 rue d'Ulm, 75005 Paris, France}
\author{R. Voituriez}
\affiliation{Laboratoire de Physique Th\'eorique de la Mati\`ere Condens\'ee, CNRS/Sorbonne Université, 
 4 Place Jussieu, 75005 Paris, France}
\affiliation{Laboratoire Jean Perrin, CNRS/Sorbonne Université, 
 4 Place Jussieu, 75005 Paris, France}
 \author{O. B\'enichou}
\affiliation{Laboratoire de Physique Th\'eorique de la Mati\`ere Condens\'ee, CNRS/Sorbonne Université, 
 4 Place Jussieu, 75005 Paris, France}
\date{\today}

\begin{abstract}
    Locally activated random walks are defined as random processes, whose dynamical parameters are modified upon visits to  given activation sites. Such  dynamics   naturally emerge in  living  systems as varied as immune  and cancer cells interacting with spatial heterogeneities in tissues, or at larger scales animals encountering local resources.  At the theoretical level, these random walks provide an explicit construction of strongly non Markovian, and aging dynamics. We propose a general analytical framework to determine  various statistical properties characterizing the position and dynamical parameters of the random walker on $d$-dimensional lattices. Our analysis applies in particular to both passive (diffusive) and active (run and tumble) dynamics, and  quantifies the aging dynamics and potential  trapping of the random walker; it finally identifies clear signatures of activated dynamics for potential use in experimental data.  
\end{abstract}

\maketitle

%\section{The locally activated lattice walker : definition and main interests}
Locally activated random walks (LARWs) are defined as stochastic processes that undergo changes in their dynamic characteristics when they encounter specific  sites, called hereafter activation sites \cite{Antal:2005,Benichou:2012,Redner:2014}. Such locally activated dynamics, where activation sites   either accelerate or slow down the process,  can be typically observed in living systems, such as  cells navigating through tissues \cite{Moreau:2018qv,Nader:2021aa}, or on a larger scale, when animals  exploit local resources, or on the contrary visit infected areas \cite{Benichou:2011b,viswanathan2011physics}.  For example, it has been observed that immune (dendritic) cells, whose function is to collect chemical signals (antigens) left by pathogens,  switch from a slow, non persistent state, to a fast and persistent state, which eventually helps triggering the specific immune response \cite{Moreau:2018qv}. This switch occurs when a threshold amount of chemical signals has been collected, after  many visits to antigen concentrated spots;  the switch   can also occur in absence of infection signals,  when the cells have been mechanically confined, as happens when they randomly migrate through tight pores distributed throughout tissues \cite{Alraies:2023aa}.

At the theoretical level, a minimal description of such dynamics suggests to endow the random walker (RWer) of interest, whose position at time $t$ will be hereafter denoted by $x(t)$, with an internal scalar variable $a(t)$ -- called activation hereafter --, which is a random variable controlled by the history of successive visits of the RWer to given activation sites up to $t$. In turn it is posited  that the parameters ruling the dynamics of the RWer, typically its instantaneous speed or persistence, depend on $a$. This minimal choice makes the dynamics of the position $x(t)$ of a LARWer genuinely non Markovian, because it depends on the past trajectory $\{x(t')\}_{t'\le t}$, and aging, because the statistics of the activation $a(t)$ is typically non stationary, eg depends on the observation time. Beyond the applications mentioned above, the class of LARWs thus provides an explicit microscopic construction of non Markovian, aging  stochastic processes with a broad range of adjustable dynamic and geometric properties, which, as we show below,  can be quantified analytically. 
Related examples of  non-Markovian RWs, in which the memory of the complete past trajectory determines the future evolution, comprise self-avoiding walks \cite{Hughes:1995}, true self-avoiding walkers \cite{PhysRevB.27.1635,PhysRevB.27.1635,PhysRevB.27.5887,PhysRevLett.119.140601}, self-interacting RWs \cite{Perman:1997,Davis:1999,Pemantle:1999,Antal:2005,Boyer:2014a,dAlessandro:2021,PhysRevX.12.011052,PhysRevLett.130.227101,Regnier:2023} and RWs with reinforcement such as the elephant walks \cite{PhysRevE.70.045101,PhysRevE.94.052134,Bercu:2021}.

So far, the analysis of LARWs has been restricted to the example of $1d$ Brownian dynamics, with a  point like activation site \cite{Benichou:2012}. Because a given point in space is almost surely never visited by a Brownian RWer for $d\ge2$, this early analysis is not suitable to   generalizations to higher space dimensions, which are of obvious importance for practical applications. In addition,  this model was limited to Brownian motion and thus parametrized by a single parameter -- the diffusion coefficient $D(a)$. It thus does not cover the case of persistent RWs -- typically parametrized by both their instantaneous speed and persistence, which are paradigmatic models of active particles, required in particular in the description of living systems as exemplified above, be them cells or animals \cite{Romanczuk:2012aa}.

% We introduce a general model of a random walker whose dynamics depends on a scalar activation level -- called activation hereafter --, which is controlled by the successive visits of the walker to given activation sites. We will mainly focus on lattice random walkers, so that activation sites are given lattice sites. We use the framework of the continuous-time random walk (\textbf{CTRW}), in which a walker at site $s$ jumps from $s$ between $t$ and $t+dt$ with probability $\frac{dt}{\tau}$, and stays at $s$ with probability $1-\frac{dt}{\tau}$. The quantity $\tau$ is henceforth referred to as the \textbf{waiting time} of the walker.
% \par
% The position of the walker at time $t$, which is the lattice site it sits on at $t$, is written $x(t)$, while the variable $a(t) = \int_0^t \mathds{1}_{\text{hotspots}}\left(x(t')\right)dt'$ is the activated time defined above. Both are stochastic quantities. \par
% In the Locally Activated Random Walk (\textbf{LARW}) model, the waiting time $\tau$ \textbf{evolves according to the activated time}. Thus, the waiting time of the walker is a \textit{stochastic} quantity. We will make explicit the activated time dependence by writing the waiting time at time $t$ as $\tau(a(t))$. This model defines a subclass of non-Markovian walkers. Indeed, the dynamics of the walker at time $t$ are strongly correlated with the walker's positions at times $t'<t$ \textit{via} the activated time. \par

In what follows, we introduce a general $d$-dimensional lattice model of continuous time LARWs, which covers the case of both simple symmetric (Polya) and persistent RWs, with either accelerated or decelerated  dynamics.  We present a general framework to obtain exact, analytical determinations of the joint distribution $P(x,a,t)$ of the position and activation of the RWer at time $t$, which fully characterizes the process. 
%Our analysis is based on an exact expression of  $P(x,a,t)$ of the non Markovian LARW in terms of the propagator $\mathcal{P}(x|0,t)$ of the corresponding {\it non-activated}, and thus Markovian RW, which is known exactly for simple and persistent RWs on $d$-dimensional lattices. 
Our analysis shows quantitatively that  activation, even if localized at a single site, can deeply impact the dynamics at large scales. For generic accelerated processes, we show that the marginal distribution of the position of the RWer (denoted $P(x,t)$ for simplicity) is non Gaussian for $d=1,2,3$ -- thereby generalizing the result of  \cite{Benichou:2012} obtained for $1d$ Brownian LARW. In contrast to the $1d$ case, which leads to anomalous diffusion, we find for $d=2,3$  a diffusive scaling $x\sim t^{1/2}$ for all choices of activation dynamics. 
%In addition, we show that $P(x,t)$  is non monotonic with the distance to the activation site $|x|$, with a local depletion around the activation site for $d=1,2$ only  \cite{Benichou:2012}, while it remains monotonic for $d=3$. 
For decelerated processes, we identify and characterize quantitatively a transition between a Gaussian, diffusive regime and a phase where the RWer can be irreversibly trapped at the activation site.

{\it LARW : general framework.}
\begin{figure}
\centering
    \begin{subfigure}{0.45\columnwidth}
		\includegraphics[width=\textwidth]{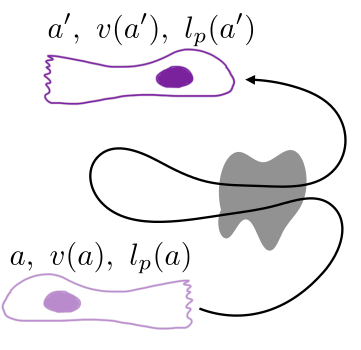}
		\vspace*{-20pt}
    \end{subfigure}
    \begin{subfigure}{0.45\columnwidth}
        \includegraphics[width=\textwidth]{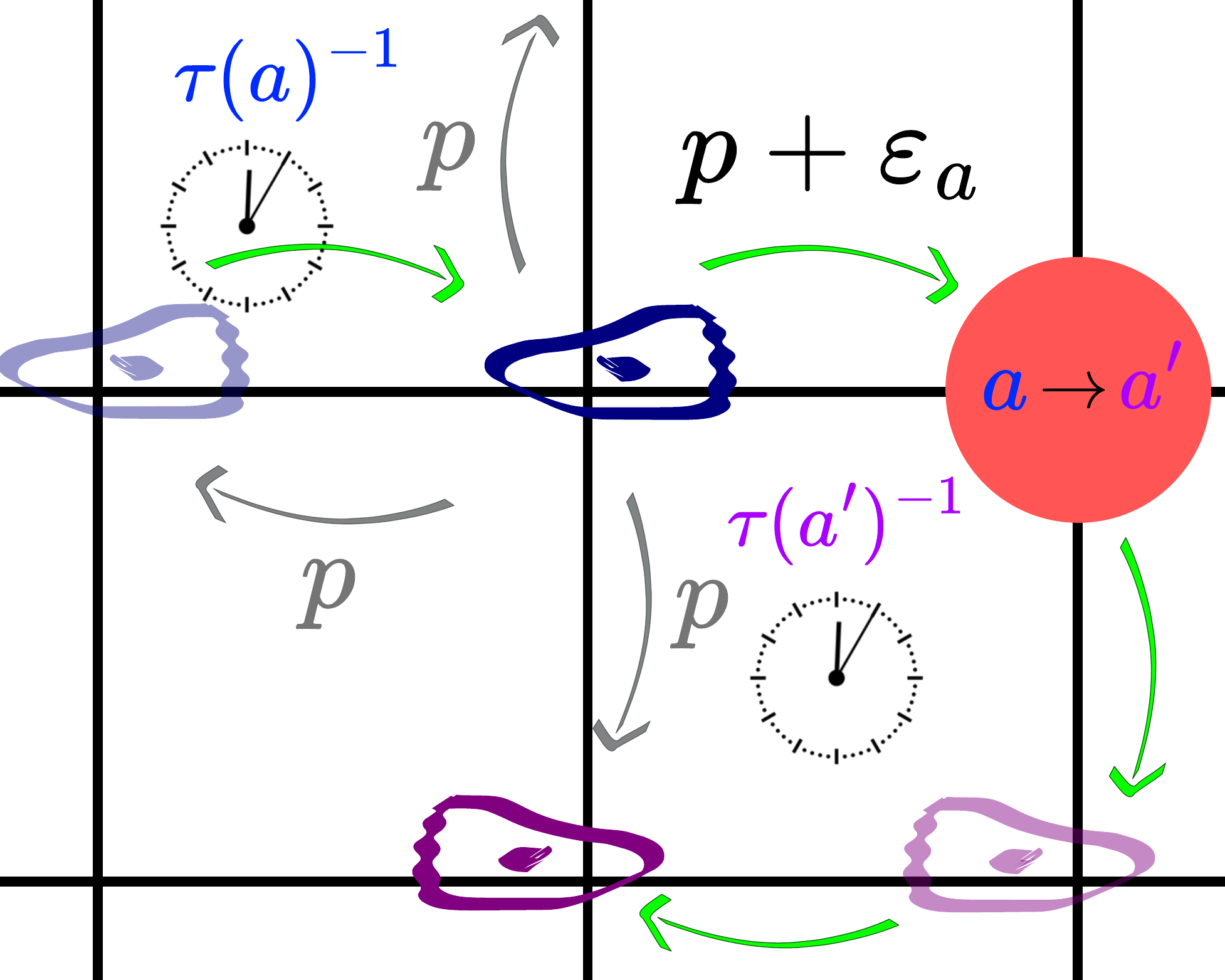}
		\vspace*{-20pt}
    \end{subfigure}
    \caption{Left: Sketch of a LARWer (eg immune cell), whose dynamic parameters (speed $v$ and persistence length $l_p$) depend on the activation time $a$, which increases upon each visit to the activation site (eg antigen carrying site). Right: A persistent LARW on a 2D square lattice, with persistence parameter $\varepsilon_a$ ($p=\frac{1-\varepsilon_a}{4}$) and waiting time $\tau(a)$. The trajectory of the RWer is represented by the green arrows, and  the activation site is represented by the  red disk. The activation time increases from $a$ to $a'$ upon the visit of the activation site. }
    \label{fig:dessin}    
\end{figure}
We  consider a RWer that   performs a generic continuous time RW on a regular $d$-dimensional infinite lattice. More precisely, the RWer, if at site $s$ with activation time $a$, performs a jump with rate $1/\tau(a)$ to a site drawn from a given distribution (see explicit examples below and Fig. \ref{fig:dessin}).  The walk starts with activation time  $a=0$ from the origin $x=0$, which is assumed by convention to be the only activation site of the lattice. In what follows, we assume that the activation $a$ of the RWer is only increasing and given by the cumulative time spent on the activation site up to time $t$, so that $a(t) = \int_0^t \mathds{1}\left(x(t')\right)dt'$, where $\mathds{1}\left(s\right)=1$ if $s$ is the activation site and 0 otherwise; in turn, $\tau(a)$ is a system dependent modelling choice, and can capture both accelerated (decreasing $\tau(a)$) and   decelerated (increasing $\tau(a)$) processes. 
Of note, the aging dynamics of $a$ makes the position process $x(t)$ non Markovian, because the jump rate of the RW depends explicitly on the activation time $a$ that is controlled by the full history of visits of the RWer to the activation site. Nonetheless, the process $(x(t),a(t))$ is Markovian, and    the joint distribution $P(x,a,t)$ fully characterizes the process ; we denote  $\hat{P}(x,a,s) = \int_0^{\infty} e^{-st} P(x,a,t) dt$ its Laplace transform. 

In what follows, we first show that even if $a(t)$ alone is not a markovian process, an explicit evolution equation for $P(0,a,t)$ can be obtained.  
Writing $P(0,t+dt,a+dt)$ as a partition  over the last time the hotspot was visited yields two different scenarii : 
(i) the walker was at $0$ at time $t$ with activation time $a$ and did not  jump during $dt$, or
(ii) the walker jumped from $0$ at an earlier time $t'<t$ with activation time $a+dt$ and came back exactly at $t+dt$.
The key point is that between two consecutive visits to the activation site, the activation time $a$, and thus the jump rate $1/\tau(a)$ of the RWer remains constant. Thus, the probability of events involved in (ii) can be written in terms of the first-passage time (FPT) density to site $0$ for a simple random  walker with constant jump rate $1/\tau(a)$ starting from a neighboring site of $0$, which is denoted $F_a(0|\partial_0, t)$. This yields (see supplementary material (SM) for details)
\begin{equation}
\begin{gathered}
    \label{single-hotspot-transport-realtime}
    \partial_t P(0,a,t) + \partial_a P(0,a,t) = \\
    -\frac{P(0,a,t)}{\tau(a)} + 
    \int_0^t \frac{dt'}{\tau(a)} P(0,a,t') F_a(0|\partial_0, t-t').
\end{gathered}
\end{equation}
Next, we define $\xi_a(s) = (1+s\tau(a))^{-1}$  and  Laplace-transform \eqref{single-hotspot-transport-realtime} to obtain a first important result
\begin{equation}
    \label{single-hotspot-transport-laplace}
    \partial_a \hat{P}(0,a,s) = \frac{[\mathcal{F}\big(0|0,\xi_a(s)\big)-1]\hat{P}(0,a,s)}{\tau(a)\xi_a(s)}
\end{equation}
where we introduced the discrete-time, non-activated first-passage generating function $\mathcal{F}(0|0,\xi)$, related to its continuous-time, activated counterpart $\hat{F}_a(0|\partial_0,s)$ by $\hat{F}_a(0|\partial_0,s) = \mathcal{F}(0|0,\xi_a(s))/\xi_a(s)$ \cite{Hughes:1995}.
Integrating this equation is straightforward and yields an explicit expression for $\hat{P}(0,a,s)$, provided that $\mathcal{F}(0|0,\xi)$ is known. \par
We now show how to obtain the full joint distribution $P(x,a,t)$ from \eqref{single-hotspot-transport-laplace}. For the walker to be on site $x$ with activation  $a$ at $t$, it must jump away from  $0$ at an earlier time $t'$ with activation $a$, and next reach site $x$ without hitting $0$ in the remaining time $t-t'$. This analysis yields the following renewal equation
\begin{equation}
\label{convolution_fulljoint}
    P(x,a,t) = \int_0^t \frac{dt'}{\tau(a)} P(0,a,t')P^a_{\text{surv}}(x|\partial_0,t-t')
\end{equation}
where we define $P^a_{\text{surv}}(x|\partial_0,t)$ to be the (survival) probability for a walker with activation $a$, starting from site $0$ and jumping at time $0^+$, to be at site $x$ at time $t$, all this without visiting site $0$ again. This quantity is related \cite{Hughes:1995} to its non-activated, discrete-time counterpart $\mathcal{P}_{\text{surv}}(x|0,\xi)$ by $\hat{P}^a_{\text{surv}}(x|\partial_0,s) = \tau(a) \xi_a(s) \frac{\mathcal{P}_{\text{surv}}(x|0,\xi_a(s))}{\xi_a(s)}$. The Laplace transform of \eqref{convolution_fulljoint} thus yields 
\begin{equation}
\label{fulljoint-laplace}
    \hat{P}(x,a,s) = \hat{P}(0,a,s) \mathcal{P}_{\text{surv}}(x|0,\xi_a(s)).
\end{equation} 
We now recall how all the quantities entering \eqref{single-hotspot-transport-laplace}, \eqref{fulljoint-laplace} can be deduced from the generating function  $\mathcal{P}(x|y,\xi)$ of the propagator of the corresponding non activated RW. Standard results \cite{Hughes:1995} yield the discrete-time generating function $\mathcal{P}_{\text{surv}}(x|0,\xi)$
\begin{equation}
\label{p_surv_lattice}
    \mathcal{P}_{\text{surv}}(x|0,\xi) = \mathcal{F}(x|0,\xi) = \frac{\mathcal{P}(x|0,\xi)}{\mathcal{P}(0|0,\xi)}
\end{equation}
as well as the first-return time to $0$
\begin{equation}
\label{fpt_discretetime}
    \mathcal{F}(0|0,\xi) = 1-\frac{1}{\mathcal{P}(0|0,\xi)}.
\end{equation}
Using these results, one finds finally the exact expression of the Laplace transformed  joint law
\begin{equation}
    \label{jointlaw_final}
    \hat{P}(x,a,s) = \frac{\mathcal{P}(x|0,\xi_a)}{\mathcal{P}(0|0,\xi_a)} \exp\left (-\int_0^a \frac{db}{\tau(b)\xi_b \mathcal{P}(0|0,\xi_b)}\right).
\end{equation}
We stress that this determination of the joint law is fully explicit for all processes for which the propagator of the underlying, non activated RW is known. We present explicit examples below.

{\it $d$-dimensional nearest neighbor LARWs.}
For the paradigmatic example of symmetric nearest neighbor RWs  on the hypercubic lattice $\mathbb{Z}^d$, one has the following integral representation of the (non activated) propagator  \cite{Hughes:1995}:
\begin{equation}
\label{propagator_hypercubic}
    \mathcal{P}\left (x = (x_k)_{k=1\dots d},\xi \right) = \int_0^\infty e^{-u} \prod_{k=1}^d I_{x_k}\left(\frac{u\xi}{d} \right) du
\end{equation}
where $I_{x_k}$ is a modified Bessel function.

For $d=1$, making use of Eq.\ref{jointlaw_final} yields (see SM):
\begin{equation}
\label{fulljoint_Z}
    \hat{P}_{1D}(x,a,s) = 
    \left(\frac{1-\sqrt{1-\xi_a^2}}{\xi_a}\right)^{|x|} \exp\left(-\int_0^a \frac{db}{\tau(b)} \frac{\sqrt{1-\xi_b^2}}{\xi_b}\right)
\end{equation}
While this expression cannot, to the best of our knowledge, be Laplace inverted analytically for arbitrary $\tau(a)$, numerical inversion is straightforward and provides the joint law at all times. Under the condition that  $\tau(a)\ll a$ for $a\to\infty$ (to be refined below), a smooth (non singular)  continuous limit, defined by  $x \to \infty, x^2\tau(a)/t$ fixed, can be obtained and yields the following asymptotic expression of the joint law of $x, a$ 
\begin{equation}
    \label{fulljoint_Z_asymp}
    P_{1D}(x,a,t) \sim \frac{Z e^{-\frac{Z^2}{2 t}}}{\sqrt{2 \pi } t^{3/2}}
\end{equation}
where $Z = \sqrt{\tau(a)}|x| + F_1(a), F_1(a) = \int_0^a \frac{db}{\sqrt{\tau(b)}}$. In turn, considering the explicit example of $\tau(a) \sim a^{-\alpha}$ with $\alpha>-1$ (to ensure $\tau(a)\ll a$ for $a\to\infty$), integration over $a$ using the saddle point method yields the marginal distribution
\begin{equation}
    \label{marginal1d}
    P_{1D}(x,t) \sim \begin{cases}
    A(\alpha) \frac{x^{\frac{1}{1+\alpha}}}{t} \exp(-B(\alpha) \frac{x^{\frac{2+\alpha}{1+\alpha}}}{4t}), \alpha>0 \\
    \frac{e^{-\frac{x^2}{2t}}}{\sqrt{2\pi t}(1-\frac{\alpha}{2} |x|)}, -1<\alpha<0
    \end{cases} 
\end{equation}
where $A,B$ are constants explicitly given in SM. 
% Due to the presence of the absolute value $|x|$, this distribution is non-Gaussian in $x$. For values of $a$ less than a threshold value $a_*(t)$, the latter being of the order of the average activated time $\langle a(t) \rangle$, we observe two nonzero maxima $\pm x_*(a,t)$. On the contrary, for $a>a_*(t)$, it is maximal only at $x_*(a,t)=0$, displaying a cusp at this point. The peculiar behavior of this joint distribution of position and local time is a signature of recurrent random walks. It is merely amplified by the activated behavior at the origin.
% Let us provide a simple explanation for this behavior in the case of a \textit{non-activated} random walker on $\mathbb{Z}$. At time $t$, the walker has spent a (local) time $a$ on site $0$ which is typically of the order $a_* \sim \sqrt{t}$. If we condition the walker to have a local time $a \gg a_*$, it will likely be close to the origin, since it has spent more time than usual there. On the contrary, if $a \ll a_*$, it is likely away from $0$. We see here a manifestation of the effects of conditioning Markov processes with respect to a certain observable \cite{Chetrite_2014}, \cite{Mazzolo_2022}. For transient random walks, the picture is different, since the number of returns to the origin is finite. After a finite time, the walker never visits the origin again; position and local time decorrelate, since the latter no longer increases. 
Of note, the marginal distribution is thus non Gaussian, and even non monotonic as a function of $x$ in the case of accelerated processes. These results generalize the earlier findings of  \cite{Benichou:2012}, where a similar $1d$ model of LARW for a Brownian particle in continuous space, with a pointlike activation site was studied. These earlier results are indeed recovered by taking the appropriate continuous limit  of  \eqref{fulljoint_Z_asymp},\eqref{marginal1d} (see SM).

\begin{figure}
\includegraphics[width=0.42\textwidth]{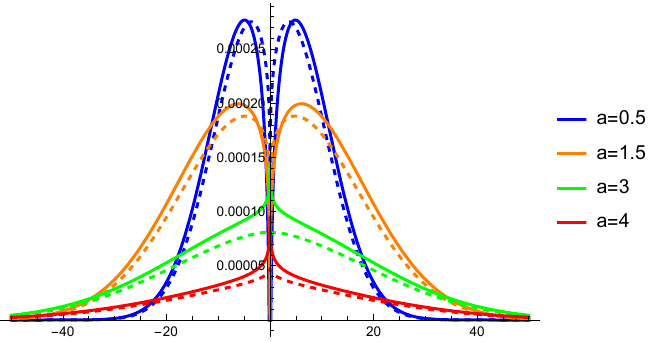}
    \caption{Joint law of activation time (colors) and position (abscissa, along a given axis) of an accelerating 2D LARW with waiting time $\tau(a) = \frac{1}{a}$. Thick lines are obtained from numerical Laplace inversion of \eqref{fulljoint_square}, while dashed lines correspond to the asymptotics \eqref{jointlaw_2d_asymp}. }
    \label{fig:theoretical_jointlaw}    
\end{figure}

We now turn to LARW on the 2--dimensional square lattice. Using \eqref{jointlaw_final},\eqref{propagator_hypercubic}, the joint law of $x=(x_1,x_2)$ and $a$ can be written in Laplace space
\begin{eqnarray}
\label{fulljoint_square}
\hat{P}_{2D}(x,a,s) = &&
\frac{\pi \int_0^\infty e^{-u} I_{x_1}\left(\frac{\xi_a u}{2}\right)I_{x_2}\left(\frac{\xi_a u}{2}\right) du}{2K(\xi_a^2)} \nonumber \\
 &&\times \exp\left (-\frac{\pi}{2}\int_0^a \frac{db}{\tau(b)\xi_bK(\xi_b^2)}\right)
\end{eqnarray}
where $K$ is the elliptic integral. While there is a priori no simple Laplace inversion of this expression,  numerical inversion can be performed and provides the joint law for all values of parameters (see Fig \ref{fig:theoretical_jointlaw}).

% In addition, making use of  \cite{hughes1995random}, the following asymptotics can be obtained:
% \begin{equation}
% \label{jointlaw_2d_laplace_asymp}
%     \hat{P}_{2D}(x,a,s) \Sim{s a \ll 1} 2K_0(2\sqrt{s\tau(a)}|x|)\frac{\exp(- \frac{\pi F_2(a)}{\log \frac{8}{\tau(a)s}})}{\log \frac{8}{\tau(a)s}}
% \end{equation} 
In addition, in the scaling regime $x \to \infty,x^2 \tau(a)/t$ fixed, Laplace inversion can  be performed analytically  (assuming $\tau(a)\ll a$ for $a\to\infty$, see SM) and yields :
\begin{equation}
\label{jointlaw_2d_asymp}
    P_{2D}(x,a,t) \sim -\frac{\tau(a)}{\pi t} e^{-\frac{|x|^2 \tau(a)}{t}} \partial_a e^{- \frac{\pi F_2(a)}{\log \frac{8 t^2}{\tau(a)^2 |x|^2}}} 
\end{equation} 
where $F_2(a) = \int_0^a \frac{db}{\tau(b)}$.
This expression of the joint law is found to be  in very good agreement with the numerical inversion (see Fig \ref{fig:theoretical_jointlaw}). Of note, different behaviours are obtained according to the value of $a$ relative to a threshold value $a_*(t)$,  which can be determined from the analysis of \eqref{jointlaw_2d_asymp}, and turns out to scale as the typical  value of $a$ at time $t$ (see SM) \footnote{A similar phenomenology is obtained for $1d$ LARW from eq. (10) }. For $a<a_*$,   trajectories with atypically low numbers of  visits to the activation site are selected. This leads to an effective repulsion from the activation site, so that   $P_{2D}(x,a,t)$ as a function of $x$ displays a local minimum for $x=(0,0)$. Conversely, for $a>a_*$,   trajectories with atypically large numbers of visits to the activation site are selected, and $P_{2D}(x,a,t)$ shows a sharp maximum for $x=(0,0)$.

%For $\tau(a) = \frac{1}{a^\alpha}$, we show in the SM that $a_*(t) \asymp \log^{\frac{1}{1+\alpha}} t$ : unsurprisingly, as underlined in \eqref{cdf_a_time}, this is the mean value of the activated time for such a choice of $\tau(a)$. 

\begin{figure}
\centering
    \begin{subfigure}{0.45\columnwidth}
		\includegraphics[width=\textwidth]{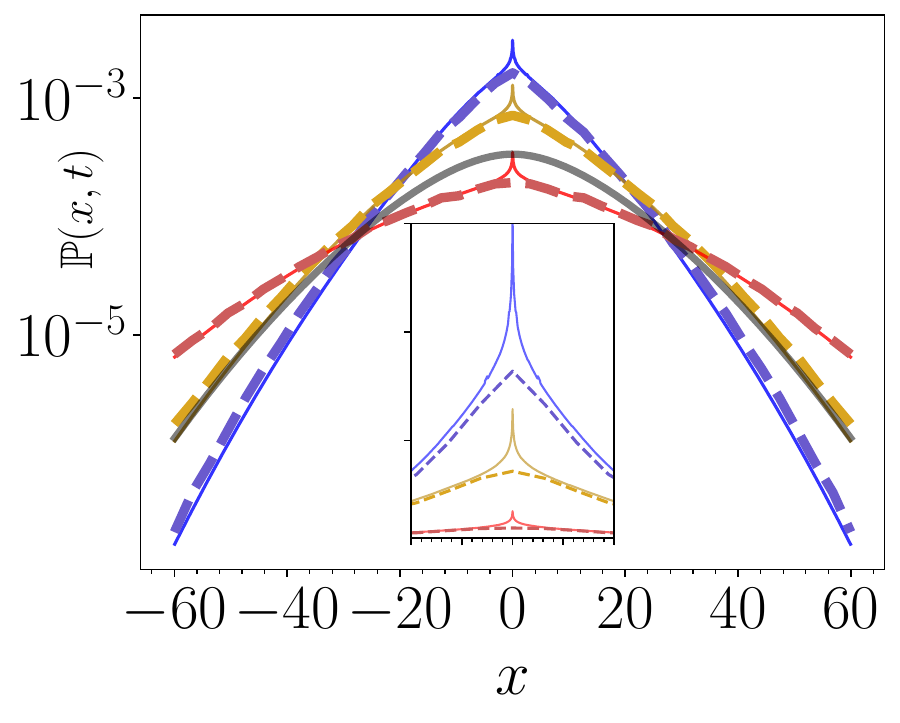}
		\begin{picture}(0,0)
		\put(-35,80){(a)}
		\end{picture}
		\label{dec_2d}
		\vspace*{-20pt}
    \end{subfigure}
    \begin{subfigure}{0.45\columnwidth}
        \includegraphics[width=\textwidth]{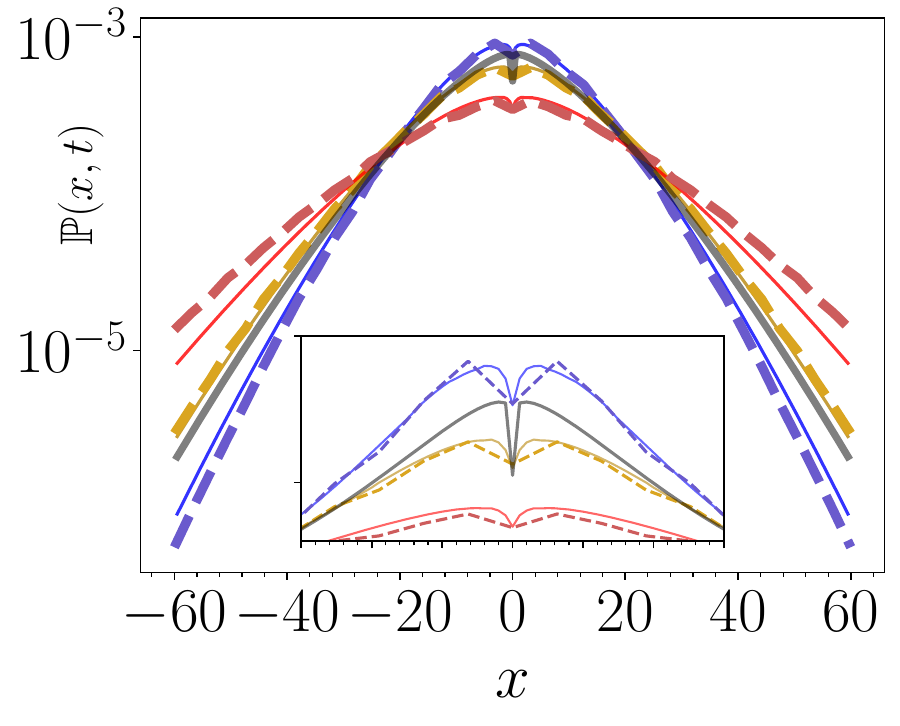}
        \begin{picture}(0,0)
		\put(-35,80){(b)}
		\end{picture}
		\vspace*{-20pt}
    \end{subfigure}
    \caption{Spatial distribution for a persistent 2D LARW with a single hotspot at the origin with persistence parameter $\varepsilon$, along a given axis $x$.  Blue, yellow and red curves correspond respectively to $\varepsilon=-0.3,0,0.4$. Black lines correspond to the saddle-point approximations \eqref{marginal2d} and are drawn for $\varepsilon=0$. Thick lines are obtained from numerical integration over $a$ of the asymptotic expression \eqref{jointlaw_2d_asymp} with rescaled waiting time $\tilde{\tau}(a) = \frac{1-\varepsilon}{1+\varepsilon}\tau(a)$, while dashed lines correspond to numerical simulations. (a) Decelerating LARW, with $\tau(a) = \sqrt{1+a}$. Note the cusp at site $x=0$. (b) Accelerating LARW, with $\tau(a) = \frac{1}{a}$. Note the nonzero typical displacement scaling logarithmically with time.}
    \label{fig:propag_2d}
\end{figure}
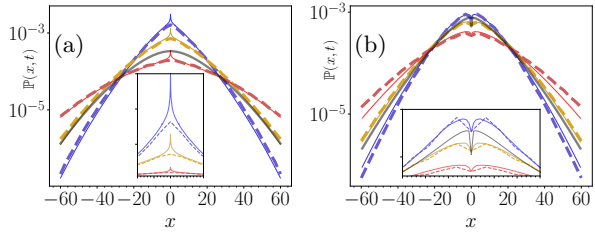

We now turn to the determination of the marginal distribution of the position at time $t$. To be explicit, we consider the example   $\tau(a) \sim \frac{1}{a^\alpha}$ with $\alpha>-1$ (to ensure $\tau(a)\ll a$ for $a\to\infty$). Using again the saddle point method in the regime $x/\sqrt{t}\gg 1$, one finds, up to subleading $\log \log$ corrections:
\begin{equation}
\label{marginal2d}
    P_{2D}(x,t) \sim \begin{cases}
    \frac{A(\alpha)}{t} (\frac{x}{\sqrt{t}})^{\frac{1-\alpha }{2 \alpha +1}} e^{-\frac{B(\alpha)(\frac{x^2}{t})^{\frac{1+\alpha}{1+2\alpha}}}{ \log^{\frac{\alpha}{1+2\alpha}} \left(C(\alpha) \left(\frac{t^{1+\alpha}}{x}\right)^{\frac{1}{1+2\alpha}}\right)}},\alpha>0 \\
    \frac{\exp(-\frac{x^2}{t})}{t \left(\pi - \frac{\alpha  x^2}{t} \log \left(\frac{8 t^2}{x^2}\right)\right)}, -1< \alpha <0
    \end{cases}
\end{equation}
where $A$ is a slowly varying function of $x$ and $t$, and $B,C$ are constants, which are explicitly given in  SM. This analytical expression of the distribution, confirmed by numerical simulations,   is strongly non-Gaussian, similarly to the $1d$ case (see Fig.\ref{fig:propag_2d}). In addition, for accelerated processes ($\alpha>0$), this distribution is maximized for  a non vanishing, increasing in time   displacement  $r_*(t)$. \par

% We obtain the distribution of the position of the LARW by integrating \eqref{fulljoint_Z} over all $a$. While it is hard to obtain an analytical expression for all values of $x$, an asymptotic estimate can be obtained from \eqref{fulljoint_Z_asymp}. An important feature is the non-Gaussian behavior of the position as soon as $\tau(a)$ varies with $a$. XXX A GARDER ?? REDITE AVEC ARTICLE BMRV XXX For an accelerating LARW with $\tau(a) = \frac{1}{a^{\alpha}}$, one finds the asymptotic result, valid in the scaling region $x \to \infty, a \ll t, \frac{x^2 \tau(a)}{t}$ fixed
% \begin{equation}
% \label{marginal_x_1d_acc}
%      P(x,t) \sim A(\alpha) \frac{x^{\frac{1}{1+\alpha}}}{t} \exp(-B(\alpha) \frac{x^{\frac{2+\alpha}{1+\alpha}}}{4t})
% \end{equation}
% where $A(\alpha), B(\alpha)$ can be computed explicitly. This distribution has two maxima at positions $\pm x_*$. We call $|x_*|$ the typical displacement of the walker. It scales as $|x_*| \sim t^{\frac{1+\alpha}{2+\alpha}}$. \par
% The LARW can also decelerate. An increasing waiting time such as $\tau(a) = (1+a)^{2\alpha}$, where $2\alpha < 1$ to keep ergodicity, yields
% \begin{equation}
% \label{marginal_x_1d_dec}
%      P(x,t) \sim \frac{e^{-\frac{x^2}{2t}}}{1+\alpha |x|} \sqrt{\frac{1}{2\pi t}}
% \end{equation}
% showing less striking non-Gaussian behavior than \eqref{marginal_x_1d_acc}. Numerically, we see that the distribution displays a cusp at $x=0$. \par 
We now turn to LARW on the $3d$ cubic lattice. As opposed to the above $d=1,2$ cases, the $3d$ non activated RW is known to be transient, so that  $P(0|0,\xi_a)$ tends to the finite value $\frac{1}{1-R}$ as $as \ll 1$, where $R\simeq 0.34...$ is the return probability on the cubic lattice \cite{Hughes:1995}. This allows for an explicit Laplace inversion in the regime $x \to \infty,a \ll t,x^2 \tau(a)/t$ fixed, which yields using \eqref{jointlaw_final} (assuming $\tau(a)\ll a$ for $a\to\infty$):
\begin{eqnarray}
\label{jointlaw_3d_asymp}
&& P_{3D}(x,a,t) \sim (1-R)\sqrt{\tau(a)} \bigg(\frac{3}{2\pi t}\bigg)^{3/2}\\
 &&\times \exp(-\frac{3|x|^2 \tau(a)}{2t} - (1-R)F_3(a)) \nonumber
\end{eqnarray}
where $F_3(a) = \int_0^a \frac{db}{\tau(b)}$. Of note, in contrast to the case of recurrent RWs for $d=1,2$,  \eqref{jointlaw_3d_asymp} is a Gaussian function of $x$ (for $a$ fixed), and is maximized at the origin, and this regardless  of $\tau(a)$, for both accelerated and decelerated processes. 
%Note that all LARW performing transient random walks, such as biased LARW, behave like the 3D LARW in this aspect, and it is possible to obtain the joint distribution of position and activated time for all transient LARW walkers by following the same derivation. \par 
Finally, taking the example of  $\tau(a) \sim 1/a^\alpha$, the marginal distribution of the position is obtained  by saddle-point integration of \eqref{jointlaw_3d_asymp} in the regime $x /\sqrt{t}\gg 1$
\begin{equation}
\label{marginal3d}
    P_{3D}(x,t) \sim 
    \begin{cases}
    \frac{A(\alpha)}{t^{3/2}} \left(\frac{x}{\sqrt{t}}\right)^{\frac{1-2\alpha}{1+2\alpha}}  e^{-B(\alpha)\left(\frac{x^2}{t} \right)^{\frac{1+\alpha}{1+2\alpha}}}, \alpha > 0 \\
    \left(\frac{3}{2\pi t}\right)^{3/2}\frac{e^{-\frac{3x^2}{2t}}}{1 - \frac{3\alpha x^2}{2t(1-R)}}, -1<\alpha < 0
    \end{cases}
\end{equation}
where $A(\alpha),B(\alpha)$ are constants and are given explicitly in the SM. Of note this distribution shows a diffusive scaling $x^2\propto t$, even if non Gaussian.  Therefore,  a localized perturbation by a single activation site, even for a transient RW that typically makes only a finite number of visits to this site, is sufficient to yield a non Gaussian behaviour.

{\it Persistent (active) LARWs.} Last, we show that \eqref{single-hotspot-transport-laplace} can in fact be used beyond the case of symmetric nearest neighbor RWs. For the sake of simplicity, we consider  a $1d$  RWer that  performs a persistent RW on $\mathbb{Z}$ \cite{springerlink:10.1007/BF01011552}, which is the discrete space analog of the classical run and tumble model of active particle. Our results below can be generalized to $d$-dimensional persistent RWs as shown in SM. In this model, after a jump $\sigma = \pm 1$, the walker performs an identical jump $\sigma$ with probability $\frac{1+\varepsilon}{2}$, and $-\sigma$ with probability $\frac{1-\varepsilon}{2}$. Local activation is taken into account by  allowing both the  jump rate $1/\tau(a)$ and the persistence parameter $\varepsilon_a \in [-1,1]$ to depend on activation $a$, whose definition is unchanged. Eq. \eqref{fulljoint-laplace} remains valid for this  process, and yields the Laplace transform of the joint distribution $\hat{P}(x\neq 0, a, s)$, given in Eq.(9) of the SM.
Importantly, this expression, upon numerical inversion, gives access to the joint law  at all times. In addition the $a s \ll 1$ asymptotics of this equation show that at large scales the persistent LARW can be   mapped exactly to   a non-persistent nearest neighbor LARW with rescaled waiting time $\tilde{\tau}(a) = \frac{1-\varepsilon(a)}{1+\varepsilon(a)} \tau(a)$, for which explicit expressions of the joint law have been obtained above. This result holds in any space dimension, %(with the prescription that for $d\ge 2$, one has $0\le\varepsilon\le 1$),
and explicitly quantifies how the activation of either persistence ($\varepsilon(a)$) or velocity ($1/\tau(a)$) impacts on the large scale dynamics of the process. In particular, acceleration  ($\tilde{\tau}(a)\to 0$) can be achieved either by increasing the instantaneous speed ($\tau(a)\to 0$) or increasing persistence ($\varepsilon(a)\to 1$), as quantified by our approach.

{\it Ergodicity breaking and trapping.} 
In the case of decelerated processes,  the particle can eventually be trapped at the activation site. Quantitatively, this trapping occurs if the particle asymptotically spends a finite fraction of time at 0, so that there is some $\gamma >0$ such that $\lim_{t\to\infty} P(0,a=\gamma t,t)>0$ and the joint law is non smooth (singular) at large times. The analysis of the joint law \eqref{jointlaw_final} shows (see SM) that such ergodicity breaking occurs if and only if  
\begin{equation}\label{trapping}
S\equiv e^{-\int_{0}^\infty\frac{da}{\tau(a)}}>0.
\end{equation}
Of note, this condition holds in all space dimensions; in the example $\tau(a)\sim a^{-\alpha}$ discussed above this occurs for  $\alpha<-1$. This condition of trapping is also realised if $\tau(a)$ diverges for a finite activation $a_f$. Importantly,  while general expressions \eqref{fulljoint_Z}, \eqref{fulljoint_square} of the joint law are valid for both trapping and non trapping regimes, explicit expressions \eqref{marginal1d}, \eqref{jointlaw_2d_asymp}, \eqref{marginal2d}, \eqref{marginal3d} have been obtained in the ergodic regime $S=0$ (or more loosely $\tau(a)\ll a$), and thus remain smooth. In the case of nearest neighbor RWs, the trapping condition \eqref{trapping} amounts to requiring that the probability $S$ that the walker remains on the activation site forever upon a given visit is non vanishing. In turn, for persistent RWs this condition applies to the effective $\tilde{\tau} $, and reveals two possible mechanisms for trapping: either waiting times diverge ($\tau(a)\to \infty)$, as in the case of trapped nearest neighbor RWs, or the RW becomes infinitely antipersistent ($\varepsilon(a)\to -1)$. 
% \begin{figure}
% \centering
%     \begin{subfigure}[T]{0.49\columnwidth}
% 		\includegraphics[width =\textwidth]{Figures/bimodal_a_cum.pdf}
% 		\begin{picture}(0,0)
% 		\put(-30,50){(a)}
% 		\end{picture}
% 		\vspace*{-20pt}
%     \end{subfigure}
%     \begin{subfigure}[T]{0.49\columnwidth}
%         \includegraphics[width=\textwidth]{Figures/bimodal_a_cum_acc_3d.pdf}
%         \begin{picture}(0,0)
% 		\put(30,80){(b)}
% 		\end{picture}
% 		\vspace*{-20pt}
%     \end{subfigure}
%     \caption{Cumulative distribution of $\frac{a(t)}{t}$ for two different LARW. (a) $1D$ antipersistent LARW, with $\tau(a) = 1, \varepsilon(a) + 1 = \frac{1}{(1+a)^{9/4}}$ so that $\tilde{\tau}(a) \gg a$ and ergodicity is broken. We can clearly see the bimodal distribution, with the rightmost bump corresponding to trajectories that are stuck in a loop of back-and-forth jumps between $0$ to $\pm 1$ : thus  $\frac{a}{t} \To{t \to \infty} \frac{1}{2}$. (b) $3D$ accelerating LARW with $\tau(a) = \frac{1}{a}$, which gets trapped with $\tau(a_f) = \infty$ for $a_f = 1.5$. The black dash-dot line represents the probability $\mathbb{P}(a(t=\infty)<a_f) = \exp(-\frac{(1-R)a_f^2}{2})$.}
%     \label{fig:bimodal}
% \end{figure}
% \par
Our analysis allows to quantify asymptotically in the large $t$ regime the dynamics of trapping by considering $P(0,a=\gamma t, t)$ in the large $t$ regime, which is obtained from the $s\to 0$ behaviour of $\hat{P}(0,a=\gamma/s , s)$ in the non-ergodic regime (see SM). This shows that for $d=1,2$, ie recurrent RWs, $P(0,a, t)\sim (1-w_d(t))\delta(a-\gamma t)$, where $\gamma =1$ for simple and persistent LARWs with $\tau(a)\to\infty$ and $\gamma = 1/2$ for persistent LARWs with $\varepsilon(a)\to -1$. The weight $w_d(t)$ is given explicitly in SM and vanishes for $t\to\infty$, so that the RWer is eventually trapped with probability 1. For $d=3$, ie transient RWs, one has $P(0,a, t)\sim e^{-(1-R) \int_0^{\infty} \frac{db}{\tau(b)}}\delta(a-\gamma t)$, and the RWer always has a non vanishing probability to remain untrapped, which is quantified by our approach.

Finally, we have presented a comprehensive analytical framework for determining a range of statistical properties that describe the dynamics of both passive and active (run and tumble) LARWs on $d$-dimensional lattices.
In the context of living systems, our analysis unveils notable and robust features of LARW (such as non-Gaussian behavior, diffusive or anomalous scaling, non-monotonicity of $P(x,t)$, and trapping). These features offer clear signatures of activated dynamics that can be potentially useful in the analysis of experimental data.

\bibliographystyle{vincent}
%\bibliographystyle{unsrt}
%\bibliography{liste}



\section*{Supplementary material}

# Supplementary Material

Julien Brémont,[1,2] Theresa Jakuszeit,[3] R. Voituriez,[1,2] and O. Bénichou[1]

[1]*Laboratoire de Physique Théorique de la Matière Condensée,*
*CNRS/Sorbonne Université, 4 Place Jussieu, 75005 Paris, France*
[2]*Laboratoire Jean Perrin, CNRS/Sorbonne Université, 4 Place Jussieu, 75005 Paris, France*
[3]*Institut Curie, CNRS UMR 144, Université PSL, 26 rue d'Ulm, 75005 Paris, France*

(Dated: November 17, 2023)

## I. PROOF OF THE MASTER EQUATION (1) FOR $P(0, a, t)$

We keep notations and definitions of the main text. Partitioning over the last time the activation site was visited yields the equation

$$P(0, a + dt, t + dt) = \mathbb{P}(\text{being at 0 with } a \text{ at } t, \text{ stay at 0 for } dt)$$
$$+ \mathbb{P}(\text{jump from 0 at } t' \leq t \text{ with } a + dt \text{ and return to 0 exactly at } t + dt)$$

$$= \left(1 - \frac{dt}{\tau(a)}\right) P(0, a, t) + \int_0^{t+dt} \frac{dt'}{\tau(a)} P(0, a + dt, t') \left(F_{a+dt}(0|\partial_0, t + dt - t')dt\right)$$

Taylor-expanding both sides of the equation and neglecting all $O(dt^2)$ terms yields (1) in the main text.

## II. ASYMPTOTIC DISTRIBUTION OF THE POSITION OF A $1D$ LARW

Here we derive equation (11) in the main text, which is the asymptotic distribution of the position of a $1D$ LARW with waiting time $\tau(a) \sim \frac{1}{a^\alpha}, \alpha > -1$.

### A. Accelerating $1D$ LARW

We assume a form $\tau(a) = \frac{1}{a^\alpha}, \alpha > 0$. We remind that the asymptotic joint distribution of $a, x$ reads (equation (10) in the main text)

$$P(x, a, t) \underset{a \ll t}{\sim} \frac{Z}{\sqrt{2\pi t^3}} e^{-Z^2/2t}, Z = a^{-\alpha/2} x + \frac{a^{1+\alpha/2}}{1 + \alpha/2} \tag{1}$$

The marginal distribution of $x$ reads, integrating over $a$ and changing variables $a = ux^{1/(1+\alpha)}$

$$P(x, t) \sim \frac{x^{1/(1+\alpha)}}{\sqrt{2\pi t^3}} \int_0^\infty x^{\frac{2+\alpha}{2+2\alpha}} \phi(u) e^{-\frac{x^{\frac{\alpha+2}{\alpha+1}}}{2t} \phi(u)^2} du, \phi(u) = \frac{u^{-\frac{\alpha}{2}} (\alpha + 2u^{\alpha+1} + 2)}{\alpha + 2} \tag{2}$$

One easily sees that $\phi$ has a minimum at $u_* = (\frac{\alpha}{2})^{1/(1+\alpha)}$, and

$$\phi(u_*) = \frac{2^{\frac{\alpha}{2\alpha+2}+1} \alpha^{-\frac{\alpha}{2\alpha+2}} (\alpha + 1)}{\alpha + 2}, \phi''(u_*) = 2^{1-\frac{3\alpha}{2\alpha+2}} \alpha^{\frac{\alpha-2}{2(\alpha+1)}} (\alpha + 1)$$

Using the Laplace method, one obtains the asymptotics for large $x$

$$P(x, t) \sim \frac{x^{\frac{4+\alpha}{2+2\alpha}}}{t} \phi(u_*) \sqrt{\frac{1}{x^{\frac{\alpha+2}{\alpha+1}} \phi''(u_*)}} e^{-\frac{x^{\frac{\alpha+2}{\alpha+1}}}{2t} \phi(u_*)^2} = \frac{A x^{\frac{1}{1+\alpha}}}{t} e^{-B x^{\frac{\alpha+2}{\alpha+1}} \frac{1}{2t}}$$

where $A = \frac{\phi(u_*)}{\sqrt{\phi''(u_*)}}, B = \phi(u_*)^2$.

### B. Decelerating $1D$ LARW

We assume here that $-1 < \alpha < 0$, so that the LARW decelerates. Also, we rather use $\tau(a) = (1 + a)^\alpha$ so that $\tau(0) = 1$ is finite. The former derivation does not apply anymore as $Z$ defined above is minimal at the boundary $a = 0$ of the integration range. It writes, near this value

$$Z^2 = |x|^2 + |x|(2 - \alpha|x|)a + O(a^2)$$

Therefore, we use the Laplace method with a maximum at the boundary to find

$$P(x,t) = \frac{1}{\sqrt{2\pi t^3}} \int_0^\infty Z e^{-\frac{Z^2}{2t}} da \underset{\sqrt{t} \ll x \ll t}{\sim} \frac{e^{-\frac{x^2}{2t}}}{\sqrt{2\pi t^3}} \int_0^\infty e^{-x(2-\alpha|x|)a/2t} |x| da = \frac{e^{-\frac{x^2}{2t}}}{\sqrt{2\pi t}(1 - \frac{\alpha}{2}|x|)}$$

## III. ASYMPTOTIC FORM OF THE JOINT LAW OF ACTIVATED TIME AND POSITION $P(x,a,t)$ FOR $2D$ LARW

### A. Laplace asymptotics of the joint law $\hat{P}(x,a,s)$

Here, we underline the derivation of the asymptotic form (14) from the exact result (13). We recall that we have, from Hughes, p151, the following behavior of the discrete-time generating function of the propagator of the square lattice $2D$ random walk, with our convention

$$P_{2D}(x,\xi) \underset{\xi \to 1^-}{\sim} 2\frac{K_0(2\sqrt{1-\xi}|x|)}{\pi}$$

When ergodicity is not broken ie $\tau(a) \ll a$, one has $\xi(s) = \frac{1}{1+s\tau(a)} \underset{as \ll 1}{\sim} 1 - s\tau(a)$.

Using the same argument, we have, for $as \ll 1$, the expansion of the term with the elliptic integral

$$K(\xi_a^2) \underset{as \ll 1}{\sim} \frac{\log \frac{8}{s\tau(a)}}{2}$$

so that

$$\frac{\pi}{2} \int_0^a \frac{db}{\tau(b)\xi_b K(\xi_b^2)} \underset{as \ll 1}{\sim} \pi \int_0^a \frac{db}{\tau(b)\log \frac{8}{s\tau(b)}} \sim \frac{\pi}{\log \frac{8}{s\tau(a)}} \int_0^a \frac{db}{\tau(b)}$$

integrating the log as if it were a constant. This is justified for algebraic $\tau(a) = a^\alpha$, for which we can perform the integral exactly and find the same $as \ll 1$ behavior.

Combining these results yields eq. (14).

### B. A result about Laplace inversion in the complex plane

We expose here a technical result that is useful for analytical inversions of Laplace transforms. Assume that we want to invert the Laplace transform $\hat{f}(s)$ where $\hat{f}(s) = o(\frac{1}{s})$ as $s \to 0$ and $\hat{f}(s)$ has no other singular point than $s = 0$. For example, one can consider $\hat{f}(s) = \frac{1}{s^\alpha}, \alpha < 1$. Also suppose $\hat{f}(\overline{z}) = \overline{\hat{f}(z)}$, and $\hat{f}$ has a branch cut along the negative real line. We write the Bromwich contour inversion of the Laplace quantity $\hat{f}(s)$ as

$$f(t) = \frac{1}{2i\pi} \int_{a-i\infty}^{a+i\infty} \hat{f}(s) e^{st} ds$$

To obtain the integral over the complex vertical line for the laplace inverse, one closes the contour from $i\infty$ to $i0^+ - \infty$ in a large circle of radius $R$, slightly hovering above the negative real line, encloses a small circle of radius $\varepsilon$ around the origin, runs below the negative real line $i0^- \to i0^- - \infty$ again and closes the contour at $-i\infty$ in a large circle. Only the contributions of the two runs along the negative real line remain as $R \to \infty$. We can then rewrite by the Cauchy residue theorem

$$f(t) = \frac{-1}{\pi} \int_0^\infty e^{-ut} \operatorname{Im} \hat{f}(-u + i0^+) du \tag{3}$$

Indeed, for example on the branch $(-\infty + i0^+, +i0^+)$ we write $s = -u + i0^+$ with $u$ running on the negative real line, and on the branch $(-\infty - i0^+, -i0^+)$ we write $s = -u - i0^+$ with $u$ running on the negative real line : we thus choose two different determinations of $\hat{f}$ due to the branch cut. Thus these two contributions yield the sum of the following two terms

$$\int_{-R}^{-\varepsilon} ds e^{st} \hat{f}(s) = \int_{\varepsilon}^{R} \hat{f}(-u + i0^+) e^{-ut} du, \quad \int_{-\varepsilon}^{-R} ds e^{st} \hat{f}(s) = -\int_{\varepsilon}^{R} \hat{f}(-u - i0^+) e^{-ut} du$$

The whole integral over the vertical complex line $f(t)$ plus these two contributions gives 0 according to the Cauchy theorem since no residue is present. This yields (3) in the $\varepsilon \to 0, R \to \infty$ limit.

## C. Application of the above to $P(x, a, t)$

As derived above, the following $as \ll 1$ asymptotics hold

$$P_{2D}(x, a, s) \sim 2K_0(2\sqrt{s\tau(a)}|x|) \frac{\exp\left(-\frac{\pi \int_0^a \frac{db}{\tau(b)}}{\log \frac{8}{\tau(a)s}}\right)}{\log \frac{8}{\tau(a)s}}$$

Let $F(a) = \int_0^a \frac{db}{\tau(b)}$. Using formula (3), one finds

$$P(\vec{x}, a, t) \underset{a \ll t}{\sim} \frac{-1}{\pi} \int_0^\infty e^{-ut} \text{Im} \left[ \frac{K_0\left(2i|x|\sqrt{u\tau(a)}\right)}{-i\pi + \log \frac{8}{u\tau(a)}} e^{-\frac{\pi F(a)}{-i\pi + \log \frac{8}{u\tau(a)}}} \right] du$$

If we change variables $v = \sqrt{\tau(a)u}$

$$P(\vec{x}, a, t) \underset{t \to \infty}{\sim} -\frac{2}{\pi^2} \text{Im} \int_0^\infty e^{-\frac{v^2 t}{\tau(a)}} v K_0\left(2i|x|v\right) \partial_a \exp\left(-\frac{\pi F(a)}{\log \frac{8}{v^2} - i\pi}\right) dv$$

Note that the main contributions to the integral are for $\frac{v^2 t}{\tau(a)}$ not too large. Hence, $v$ should scale as $\sqrt{\frac{\tau(a)}{t}}$. For $x\sqrt{\frac{\tau(a)}{t}} \gg 1$, one can use the asymptotic form $K_0(2i|x|v) \sim \frac{\sqrt{\pi}e^{-2iv|x|}}{2\sqrt{iv|x|}}$. One then finds

$$P(\vec{x}, a, t) \underset{t \to \infty, x\sqrt{\frac{\tau(a)}{t}} \gg 1}{\sim} -\frac{1}{\pi^{3/2}\sqrt{|x|}} \text{Im} \frac{1}{\sqrt{i}} \int_0^\infty e^{-\frac{v^2 t}{\tau(a)} - 2iv|x|} \sqrt{v} \partial_a \exp\left(-\frac{\pi F(a)}{\log \frac{8}{v^2} - i\pi}\right) dv$$

Completing the square in the exponential, we write $u = v\sqrt{\frac{t}{\tau(a)}} + i|x|\sqrt{\frac{\tau(a)}{t}}$

$$P(\vec{x}, a, t) \underset{t \to \infty}{\sim} -\frac{1}{\pi^{3/2}\sqrt{|x|}} e^{-\frac{|x|^2 \tau(a)}{t}} \sqrt{\frac{\tau(a)}{t}} \text{Im} \frac{1}{\sqrt{i}} \int_{i|x|\sqrt{\frac{\tau(a)}{t}}}^{\infty + i|x|\sqrt{\frac{\tau(a)}{t}}} e^{-u^2} \sqrt{v(u)} \partial_a \exp\left(-\frac{\pi F(a)}{\log \frac{8}{v(u)^2} - i\pi}\right) du$$

This integral is dominated by small values of $u$ : therefore, $v^2 \sim -\frac{\tau(a)^2 |x|^2}{t^2} = \frac{\tau(a)^2 |x|^2}{t^2} e^{-i\pi}$, because we approach this value from the lower half plane. This simplifies the $-i\pi$ in the denominator of the exp factor, and we are left with a simple gaussian integral. This finally yields for the joint law

$$P(\vec{x}, a, t) \underset{t \to \infty, \sqrt{t} \ll x \ll t}{\sim} -\frac{\tau(a)}{\pi t} e^{-\frac{|x|^2 \tau(a)}{t}} \partial_a e^{-\frac{\pi F(a)}{\log \frac{8t^2}{\tau(a)^2 |x|^2}}} \tag{4}$$

Note that by expanding the Bessel $K_0$ further in $1/x$, we can have a better estimate.

### D. Determining the threshold $a_*(t)$

In the main text, we claim that above a threshold activated time $a_*(t)$, the joint distribution $P(x,a,t)$ displays a cusp at $x=0$, while it displays bumps for $a < a_*(t)$. We perform the analysis for the choice $\tau(a) = \frac{1}{a^q}$ and compute this threshold in this section. We will show that

$$a_*(t) = \log^{\frac{1}{1+q}} t$$

To do so, we look for values of $a$ such that $\partial_x P(x,a,t) = 0$ has a solution at $x \neq 0$. Looking at the structure of $P(x,a,t)$ in $x$, it suffices to know for which values of $a$ one has $\partial_x P(x,a,t)_{|x=1} < 0$, since it changes sign for large $x$.

We expect to find $a_*(t) \asymp \log^\delta t$. Writing $t = e^u$ and computing the $x$ derivative at $x=1$ of the joint law, and after some algebra, we have at largest order in $u$

$$\partial_x P(x,a,t)_{|x=1} < 0 \iff \pi u^{1+\delta(2q+1)} - 4(1+q)^2 u^{2+q\delta} < 0 \text{ for } u \text{ large enough} \iff \delta < \frac{1}{1+q}$$

Therefore, the threshold value is

$$a_*(t) = \log^{\frac{1}{1+q}} t$$

which is of the order of the mean of the activated time at time $t$, as expected : this mean sets a natural scale for $a$.

## IV. ASYMPTOTIC DISTRIBUTION OF THE POSITION OF A $2D$ LARW

In this section we derive equation (14) in the main text, which is the asymptotic distribution of the position of a $2D$ LARW with waiting time $\tau(a) = \frac{1}{a^\alpha}, \alpha > -1$.

### A. Decelerating $2D$ LARW

The reasoning is the same as for the $1D$ LARW.

### B. Accelerating $2D$ LARW

First, we rewrite the asymptotic joint law $P(x,a,t)$ in the more computationally practical manner

$$P(x,a,t) \sim -\frac{\tau(a)}{\pi t} e^{-\frac{|x|^2 \tau(a)}{t}} \partial_a e^{-\frac{\pi F(a)}{\log \frac{8t^2}{\tau(a)^2 |x|^2}}} = \frac{\left((\alpha+1)\log\left(\frac{2\sqrt{2}ta^\alpha}{x}\right) - \alpha\right)\exp\left(-\frac{x^2 a^{-\alpha}}{t} - \frac{\pi a^{\alpha+1}}{(\alpha+1)\left(2\log\left(\frac{2\sqrt{2}ta^\alpha}{x}\right)\right)}\right)}{2(\alpha+1)t\log^2\left(\frac{2\sqrt{2}ta^\alpha}{x}\right)}$$

We follow the same derivation as the $3D$ case. Doing so, we obtain the following Laplace-type integral, after the change of variables $a = ux^{\frac{2}{1+2\alpha}}$

$$P_{2D}(x,t) \sim x^{\frac{2}{1+2\alpha}} \int_0^\infty du g(x,t,u) e^{-x^{\frac{2+2\alpha}{1+2\alpha}} \phi(u)} \tag{5}$$

where the large-deviation function writes

$$\phi(u) = \frac{\pi u^{\alpha+1}}{2(\alpha+1)\log\left(\frac{2\sqrt{2}t\left(ux^{\frac{2}{2\alpha+1}}\right)^\alpha}{x}\right)} + \frac{1}{tu^\alpha}$$

and the algebraic prefactor in the integrand writes

$$g(x,t,u) = \frac{(\alpha+1)\log\left(\frac{2\sqrt{2}t\left(ux^{\frac{2}{2\alpha+1}}\right)^{\alpha}}{x}\right) - \alpha}{2(\alpha+1)t\log^2\left(\frac{2\sqrt{2}t\left(ux^{\frac{2}{2\alpha+1}}\right)^{\alpha}}{x}\right)}$$

$\phi$ has a local minimum at a value $u_*(t)$ that verifies the equation

$$\phi'(u_*) = 0 \iff \frac{\pi u_*^{\alpha}}{2\log\left(\frac{2\sqrt{2}t\left(ux^{\frac{2}{2\alpha+1}}\right)^{\alpha}}{x}\right)}\left(1 - \frac{\alpha}{(1+\alpha)\log\left(\frac{2\sqrt{2}t\left(u_*x^{\frac{2}{2\alpha+1}}\right)^{\alpha}}{x}\right)}\right) - \frac{u_*^{1-\alpha}\alpha}{t} = 0$$

There is no simple analytical form for $u_*$. To proceed further, we perform a perturbative expansion in $\log\left(\frac{2\sqrt{2}t\left(u_*x^{\frac{2}{2\alpha+1}}\right)^{\alpha}}{x}\right)$, which, for large $tx^{\frac{-1}{2\alpha+1}}$, should be large enough. Since $x \ll t$ and $\alpha > 0$, $tx^{\frac{-1}{2\alpha+1}} \gg 1$ is satisfied. Of course, since the log varies slowly, we expect that traces of this perturbative expansion will remain in the comparison with numerics. This leads to the simpler implicit equation

$$u_* \sim \left(\frac{2\alpha}{\pi t}\log\left(\frac{2\sqrt{2}t\left(u_*x^{\frac{2}{2\alpha+1}}\right)^{\alpha}}{x}\right)\right)^{\frac{1}{1+2\alpha}}$$

We now plug this solution in (5) and perform the saddle-point approximation, while only keeping log terms and neglecting the log log terms. This yields, after some algebra performed with Mathematica, the following saddle-point form of the propagator announced in the main text

$$P(x,t) \sim \frac{A(\alpha)}{t}(\frac{x}{\sqrt{t}})^{\frac{1-\alpha}{2\alpha+1}}\exp\left(-B(\alpha)\frac{(\frac{x^2}{t})^{\frac{1+\alpha}{1+2\alpha}}}{l^{\frac{\alpha}{1+2\alpha}}}\right) \tag{6}$$

where

$$l = \log\left(2\sqrt{2}\left(\frac{\left(\frac{2}{\pi}\right)^{\alpha}\alpha^{\alpha}t^{\alpha+1}}{x}\right)^{\frac{1}{2\alpha+1}}\right), B(\alpha) = \frac{\left(\frac{\pi}{2\alpha}\right)^{\frac{\alpha}{2\alpha+1}}(2\alpha+1)}{\alpha+1}, A(\alpha) = \frac{\sqrt{\frac{\left(\frac{2}{\pi}\right)^{\frac{1-\alpha}{2\alpha+1}}\alpha^{\frac{\alpha+2}{2\alpha+1}}}{2\alpha+1}}\left(l - \frac{\alpha}{\alpha+1}\right)}{l^{\frac{7\alpha+2}{4\alpha+2}}}$$

The term $C(\alpha)$ used in the main text, which is defined by $l = \log\left(C(\alpha)\left(\frac{t^{1+\alpha}}{x}\right)^{\frac{1}{1+2\alpha}}\right)$ thus writes $C(\alpha) = 2\sqrt{2}\left(\left(\frac{2}{\pi}\right)^{\alpha}\alpha^{\alpha}\right)^{\frac{1}{2\alpha+1}}$.

## V.  3D LARW

In this section we derive equation (16) in the main text, which is the asymptotic distribution of the position of a $3D$ LARW with waiting time $\tau(a) = \frac{1}{a^{\alpha}}, \alpha > -1$. In [1] is given the generating function of the propagator of a RW on the cubic lattice

$$P_{3D}(x,\xi) \sim 3\frac{\exp\left(-\sqrt{6(1-\xi)}|x|\right)}{2\pi|x|}$$

### A. Decelerating 3D LARW

We use eq. (12) in the main text, and integrate it over all $a$ to obtain the marginal distribution of $x$. We first focus on the decelerating case so that $\tau(a) = (1+a)^\alpha, 0 < \alpha < 1$. We write

$$P_{3D}(x,t) = \int_0^\infty da P_{3D}(x,a,t) \sim (1-R)\left(\frac{3}{2\pi t}\right)^{3/2} \int_0^\infty da \sqrt{\tau(a)} \exp\left(-\frac{3|x|^2 \tau(a)}{2t} - (1-R)\int_0^a \frac{db}{\tau(b)}\right)$$

In the scaling region $\frac{x^2}{t} \sim 1$, this integral is dominated by values of $a$ close to $a_* = 0$. We then apply the Laplace method with a function exhibiting a boundary extremum. The series expansion of the term in the exponential at $a = 0$ is, after computing the integral $\int_0^a \frac{db}{\tau(b)} = \frac{\left((a+1)^{1-\alpha}-1\right)}{1-\alpha}$

$$\frac{(1-R)\left((a+1)^{1-\alpha}-1\right)}{1-\alpha} + \frac{(3x^2)(a+1)^\alpha}{2t} = \frac{3x^2}{2t} + a\left(1 - R + \frac{\alpha(3x^2)}{2t}\right) + O\left(a^2\right)$$

Therefore, the Laplace method yields the asymptotic form

$$P_{3D}(x,t) \sim \left(\frac{3}{2\pi t}\right)^{3/2} \frac{e^{-\frac{3x^2}{2t}}}{1 + \frac{3\alpha x^2}{2t(1-R)}}$$

as written in the main text.

### B. Accelerating 3D LARW

We now focus on $\tau(a) = \frac{1}{a^\alpha}, \alpha > 0$. Following the same derivation as above, one has

$$P_{3D}(x,t) \sim (1-R)\left(\frac{3}{2\pi t}\right)^{3/2} \int_0^\infty da \sqrt{\frac{1}{a^\alpha}} \exp\left(-\frac{3|x|^2}{2ta^\alpha} - (1-R)\frac{a^{1+\alpha}}{1+\alpha}\right)$$

We change variables $a = ux^{\frac{2}{1+2\alpha}}$ so that

$$P_{3D}(x,t) \sim (1-R)\left(\frac{3}{2\pi t}\right)^{3/2} x^{\frac{2}{1+2\alpha}} \int_0^\infty du \sqrt{\frac{1}{u^\alpha x^{\frac{2\alpha}{1+2\alpha}}}} e^{-x^{\frac{2+2\alpha}{1+2\alpha}}\phi(u)}$$

where $\phi(u) = \frac{(1-R)u^{\alpha+1}}{\alpha+1} + \frac{3}{2tu^\alpha}$ is the large deviation function associated to this Laplace integral. It is minimal at

$$u_*(t) = 3^{\frac{1}{2\alpha+1}}\left(\frac{\alpha}{2t-2Rt}\right)^{\frac{1}{2\alpha+1}}, \phi(u_*(t)) = \frac{3(2\alpha+1)\left(3^{\frac{1}{2\alpha+1}}\left(\frac{\alpha}{2-2R}\right)^{\frac{1}{2\alpha+1}}\right)^{-\alpha}}{2(\alpha+1)t^{\frac{1+\alpha}{1+2\alpha}}}$$

The Laplace method yields, after some algebra, in the regime $\left(\frac{x^2}{t}\right)^{\frac{1+\alpha}{1+2\alpha}} \gg 1$

$$P(x,t) \sim \frac{A(\alpha)}{t^{3/2}}\left(\frac{x}{\sqrt{t}}\right)^{\frac{1-2\alpha}{1+2\alpha}} e^{-B(\alpha)\left(\frac{x^2}{t}\right)^{\frac{1+\alpha}{1+2\alpha}}}$$

where

$$A(\alpha) = \frac{3^{\frac{1}{2\alpha+1}+1}(1-R)\left(\frac{\alpha}{2(1-R)}\right)^{\frac{1}{2\alpha+1}}}{\sqrt{2\pi}\sqrt{\alpha(2\alpha+1)}}, B(\alpha) = \frac{3(2\alpha+1)\left(3^{\frac{1}{2\alpha+1}}\left(\frac{\alpha}{2(1-R)}\right)^{\frac{1}{2\alpha+1}}\right)^{-\alpha}}{2(\alpha+1)}$$

## VI. PERSISTENT (ACTIVE) LARW

### A. Exact determination of the propagator of $1D$ persistent random walks

Here, we derive the Laplace transform of the joint distribution of $x, a$ for persistent $1D$ LARWs. Consider a persistent *non-activated* walker on $\mathbb{Z}$. The probability for the walker to jump in the same direction as its last step is $\frac{1+\varepsilon}{2}$. In [2], this corresponds to the so-called 'forward jump model'. In this article, the lattice Green function is given in Fourier space and writes, in Fourier space

$$P(q,\xi) = \sum_{n=0}^{\infty} \xi^n \sum_{x \in \mathbb{Z}} e^{iqx} P(x,n) dx = \frac{1 - \varepsilon \xi \cos q}{1 + \varepsilon \xi^2 - \xi(1+\varepsilon)\cos q} = \frac{1}{1+\varepsilon\xi^2} \frac{1 - \varepsilon\xi\cos q}{1 - \alpha\cos q}, \alpha = \frac{\xi(1+\varepsilon)}{1+\varepsilon\xi^2}$$

Writing $\frac{1}{1-\alpha\cos q} = \int_0^\infty du e^{-u(1-\alpha\cos q)}$, and using the trigonometric identity $\cos nq \cos q = \frac{\cos((n+1)q)+\cos((n-1)q)}{2}$, we obtain an integral representation in terms of modified Bessel functions $I_n, I_{n\pm 1}$ for the inverse Fourier transform $q \to x$ of $P(q,\xi)$:

$$P(x = n,\xi) = \frac{1}{2\pi}\int_{-\pi}^{\pi} P(q,\xi)\cos(nq) = \frac{1}{1+\varepsilon\xi^2}\int_0^{+\infty} du e^{-u}\left[I_n(\alpha u) - \frac{\varepsilon\xi}{2}\big(I_{n+1}(\alpha u) + I_{n-1}(\alpha u)\big)\right] \tag{7}$$

This yields, after some algebra performed in Mathematica, the generating function of the propagator for the persistent RW for $x \neq 0$ (the propagator has a different expression for $x = 0$ and $x \neq 0$, whenever $\varepsilon \neq 0$)

$$\frac{P(x,\xi)}{P(0,\xi)} = \frac{\left(\frac{\alpha}{\sqrt{1-\alpha^2}+1}\right)^{|x|}(\xi_a\varepsilon_a - \alpha)}{\left(1 - \sqrt{1-\alpha^2}\right)\xi_a\varepsilon_a - \alpha}, P(0,\xi) = \frac{1-\varepsilon}{(1-\xi^2)\varepsilon - \sqrt{(1-\xi^2)(1-\xi^2\varepsilon^2)}}, \alpha = \frac{\xi(1+\varepsilon)}{1+\varepsilon\xi^2} \tag{8}$$

We can check that it reduces to the symmetric random walk (Polya) propagator for $\varepsilon = 0$. Making use of Eq. (7) of the main text, one obtains explicitly the Laplace tranform of the joint distribution:

$$\hat{P}(x \neq 0, a, s) = \frac{\left(\frac{1-\sqrt{1-\alpha^2}}{\alpha}\right)^{|x|}(\xi_a\varepsilon_a - \alpha)}{\left(1 - \sqrt{1-\alpha^2}\right)\xi_a\varepsilon_a - \alpha}\exp\left(-\int_0^a \frac{db}{\tau(b)\xi_b}\frac{\sqrt{(1-\xi_b^2)(1-\varepsilon_b^2\xi_b^2)} - \varepsilon_b(1-\xi_b^2)}{(1-\varepsilon_b)}\right), \alpha = \frac{\xi_a(\varepsilon_a+1)}{\xi_a^2\varepsilon_a+1} \tag{9}$$

### B. Small-$s$ asymptotics of the $1D$ persistent LARW

In the case where $\tau(a) \ll a$, (8) can be expanded for $as \ll 1$ to yield the long-time limit, for $x \neq 0$

$$\hat{P}(x,a,s) \underset{as\ll 1}{\sim} \exp\left(-[x(1-\varepsilon(a)) + \varepsilon(a)]\sqrt{\frac{2s\tau(a)}{(1+\varepsilon(a))(1-\varepsilon(a))}} - \sqrt{2s}\int_0^a da'\sqrt{\frac{1+\varepsilon(a')}{\tau(a')(1-\varepsilon(a'))}}\right) \tag{10}$$

Or, using the rescaled waiting time $\tilde{\tau}(a) = \frac{1-\varepsilon(a)}{1+\varepsilon(a)}\tau(a)$

$$\hat{P}(x,a,s) \underset{as\ll 1}{\sim} \exp\left(-x\sqrt{2s\tilde{\tau}(a)} - \frac{\varepsilon(a)}{1-\varepsilon(a)}\sqrt{2s\tilde{\tau}(a)} - \sqrt{2s}\int_0^a da'\frac{1}{\sqrt{\tilde{\tau}(a')}}\right) \tag{11}$$

Laplace inverting the above, one obtains

$$P(x,a,t) \underset{a\ll t}{\sim} \frac{\sqrt{\tilde{\tau}(a)}(x + \frac{\varepsilon(a)}{1-\varepsilon(a)}) + \int_0^a da'\frac{1}{\sqrt{\tilde{\tau}(a')}}}{\sqrt{2\pi t^3}}\exp\left(-\frac{\left(\sqrt{\tilde{\tau}(a)}(x + \frac{\varepsilon(a)}{1-\varepsilon(a)}) + \int_0^a da'\frac{1}{\sqrt{\tilde{\tau}(a')}}\right)^2}{2t}\right) \tag{12}$$

Except when $\varepsilon(a) \to 1^-$, which corresponds to a diverging persistence length $L_p(a) = \frac{1}{1-\varepsilon(a)}$, the above, in the limit $a \ll t, x \gg \varepsilon(a)L_p(a)$, indeed tends to the joint distribution of $x, a$ for a non-persistent LARW with rescaled waiting time $\tilde{\tau}(a)$.

In the particular case of a diverging persistence length, we write $\varepsilon(a) = 1 - \frac{1}{a^\alpha}, \alpha > 0$ and study the case $\tau(a) = 1$ for simplicity, so that only persistence varies with $a$. Thus $\tilde{\tau}(a) \sim 2a^{-\alpha}$, and (12) becomes

$$P(x,a,t) \underset{1 \ll a \ll t}{\sim} \frac{\sqrt{2}(xa^{-\alpha/2} + a^{\alpha/2}) + \frac{a^{1+\alpha/2}}{(1+\alpha/2)\sqrt{2}}}{\sqrt{2\pi t^3}} \exp\left(-\frac{\left(\sqrt{2}(xa^{-\alpha/2} + a^{\alpha/2}) + \frac{a^{1+\alpha/2}}{(1+\alpha/2)\sqrt{2}}\right)^2}{2t}\right) \quad (13)$$

For $x \gg L_p(a) = a^\alpha$, which corresponds to length scales $x$ much larger than the persistence length, we get back to the behavior of the non-persistent LARW. However, for $x \ll a^\alpha$ this equivalence does not hold. We now show that this does not impact the marginal distribution of $x$. We integrate Eq.((13)) over $a$ and note that the minimum of the function $a \mapsto xa^{-\alpha/2} + a^{\alpha/2} + \frac{a^{1+\alpha/2}}{(1+\alpha/2)\sqrt{2}}$, for large $x$, is realized for a value $a_m \propto x^{\frac{1}{1+\alpha}}$. Define

$$\phi(u,x) = \frac{\sqrt{2}(2 + \alpha + u^{1+\alpha})}{u^{\alpha/2}(2 + \alpha)}$$

so that after a change of variables $a = ux^{\frac{1}{1+\alpha}}$, with $u = O(1)$, one has for large $x$

$$P(x,t) \sim \frac{x^{1/(1+\alpha)}}{\sqrt{2\pi t^3}} \int_0^\infty x^{\frac{\alpha+2}{2+2\alpha}} \phi(u,x) e^{-\phi(u,x)^2 \frac{\alpha+2}{2t}} du$$

The argminimum $u_*(x)$ of $\phi(u,x)$ is $u_*(x) \underset{x \to \infty}{\sim} \alpha^{1/(1+\alpha)} + o(1)$, which is indeed $O(1)$ with respect to $x$. Hence,

$$\phi(u_*) \sim \frac{2\sqrt{2}\alpha^{-\frac{\alpha}{2(\alpha+1)}}(\alpha+1)}{\alpha+2}, \phi''(u_*(x)) \sim \frac{\alpha^{\frac{\alpha-2}{2(\alpha+1)}}(\alpha+1)}{\sqrt{2}}$$

By the Laplace method, one deduces the asymptotic form of $P(x,t)$ for large $x$

$$P(x,t) \sim \frac{Ax^{\frac{1}{1+\alpha}}}{t} e^{-B\frac{x^{\frac{\alpha+2}{\alpha+1}}}{2t}}$$

This is is the same behavior as the non-persistent walker, with the same constants. The key to understand this result is that the typical persistent length of walkers at position $x$ is $L_p^*(x) \sim (u_*(x)x^{\frac{1}{1+\alpha}})^\alpha \ll x$, so that the diffusive regime is attained as long as $x$ is sufficiently large. Thus, the divergence of the persistence length asymptotically impacts the joint law of $a$ and $x$, but not the marginal law of $x$.

### C. Ergodicity breaking (loops) for antipersistent 1D LARW

Consider a 1D, persistent LARW, with $\tau(a) = \frac{1}{a^\beta}$ and $\varepsilon(a) = -1 + \frac{1}{a^\gamma}$ so that $\tilde{\tau}(a) = \frac{1-\varepsilon(a)}{1+\varepsilon(a)}\tau(a) \sim 2a^{\gamma-\beta}$. Let $k$ denote the number of returns to the activation site, and $a_k$ denote the typical scale of the activated time for a walker which has returned $k$ times to the activation site. Then $a_{k+1} \sim a_k + \tau(a) = a_k + \frac{1}{a^\beta}$, so that $a_k \propto k^{\frac{1}{1+\beta}}$. There is a finite probability that the walker is stuck in an antipersistent loop between $0$ and $\pm 1$ if $\prod_k \frac{1-\varepsilon(a_k)}{2} > 0$, or, equivalently, if $\sum \frac{1}{a_k^\gamma} < \infty$, which means that $\gamma > 1 + \beta$ since $a_k^\gamma \sim k^{\frac{\gamma}{1+\beta}}$. This means that $\tilde{\tau}(a) = \frac{1-\varepsilon_a}{1+\varepsilon_a}\tau(a) \gg a$. This is the same criterion as for the nonpersistent walker, only with the appropriately rescaled waiting time $\tilde{\tau}$.

### D. Persistent LARW for $d \geq 2$

The propagator of discrete time, non activated persistent lattice random walks has been obtain in Fourier space in [2] for $d = 2, 3$. The above analysis of $1d$ persistent LARW can thus in principle be generalized to $d = 2, 3$. We expect the equivalence to rescaled diffusive LARW to hold as for $d = 1$. This analysis involves tedious algebra that we leave for further works ; we however confirm this equivalence numerically for $d = 2$ in Fig. (1).

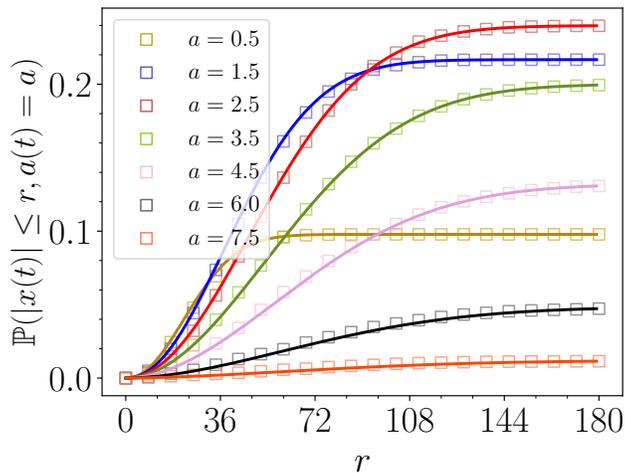

FIG. 1: Cumulative distribution of the radius $r = |x(t)|$, at given activated time $a$, of an accelerating anti-persistent 2D LARW with waiting time $\tau(a) = \frac{1}{a}$ and persistence parameter $\varepsilon = -0.3$. Solid lines are obtained from numerical Laplace inversion of the joint distribution for a *non-persistent* LARW, eq. (15) of the main text, replacing $\tau(a)$ by the effective waiting time $\tilde{\tau}(a) = \frac{1-\varepsilon}{1+\varepsilon} \tau(a)$. Symbols are obtained from numerical simulations.

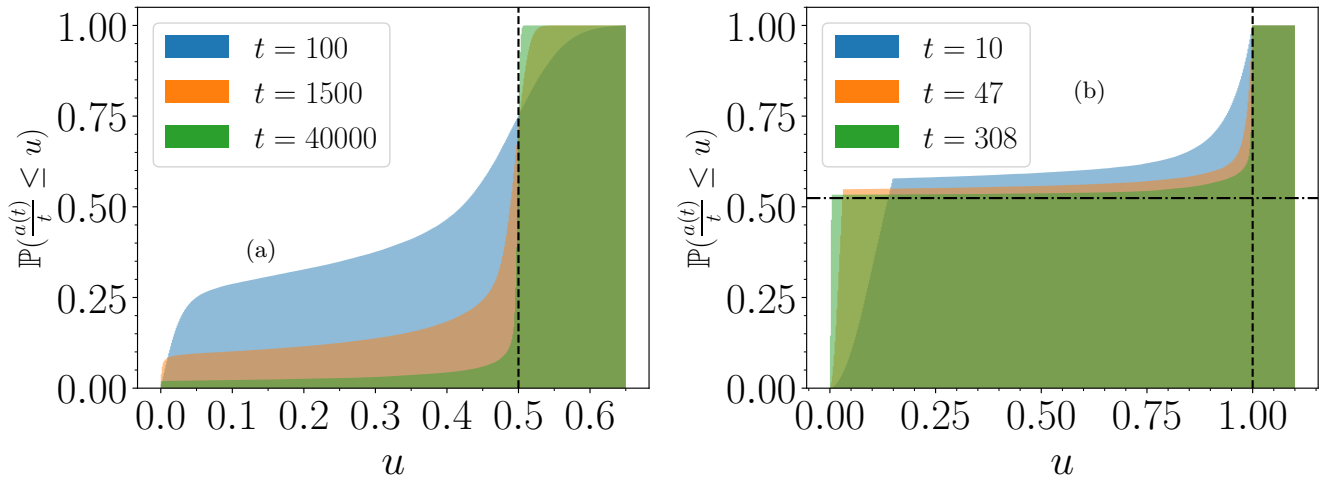

FIG. 2: Cumulative distribution of the $\frac{a(t)}{t}$ for two different LARW. (a) $1D$ antipersistent LARW, with $\tau(a) = 1, \varepsilon(a) + 1 = \frac{1}{(1+a)^{9/4}}$ so that $\tilde{\tau}(a) \gg a$ and ergodicity is broken. We can clearly see the bimodal distribution, with the rightmost bump corresponding to trajectories that are stuck in a loop of back-and-forth jumps between 0 to $\pm 1$ : thus $\frac{a}{t} \underset{t \to \infty}{\to} \frac{1}{2}$. (b) 3D accelerating LARW with $\tau(a) = \frac{1}{a}$, which gets trapped with $\tau(a_f) = \infty$ for $a_f = 1.5$. The black dash-dot line represents the probability $\mathbb{P}(a(t = \infty) < a_f) = \exp\left(-\frac{(1-R)a_f^2}{2}\right)$.

## VII. LONG TIME ASYMPTOTICS OF $a(t)$ IN THE ERGODICITY BREAKING REGIME

### A. General method of obtaining asymptotics in the ergodicity breaking regime

As underlined in the main text, in the regime where $S = e^{-\int_0^\infty \frac{db}{\tau_b}} > 0$, we can no longer use the asymptotic forms $as \ll 1$ that relied on the fact that $a(t) \ll t$, since the walker can now get trapped forever. We have to carefully perform our asymptotic expansions to obtain a scaling form of the trapping probability $P(0, a = \gamma t, t)$. To do so, we first recall the (exact) expression from the main text

$$\hat{P}(0, a, s) = \exp\left(-\int_0^a \frac{db}{\xi_b \tau_b P(0|0, \xi_b)}\right) \tag{14}$$

and we are now interested in the quantity $\hat{P}(0, a, s)$ where $as \sim 1$. The integral $\int_0^a \frac{db}{\xi_b \tau_b P(0|0,\xi_b)}$ contains two different regimes, depending on whether $s\tau_b \gg 1$ or $s\tau_b \ll 1$. Assuming $b \mapsto \tau_b$ is monotonically strictly increasing, these are separated by the value $b_* = \tau^{-1}(1/s)$. Since, by assumption, $\tau_b \gg b$, we have $b_* a \ll 1$ and we can rewrite the integral as

$$\int_0^a \frac{db}{\xi_b \tau_b P(0|0,\xi_b)} = \int_0^{\tau^{-1}(1/s)} \frac{db}{\xi_b \tau_b P(0|0,\xi_b)} + \int_{\tau^{-1}(1/s)}^a \frac{db}{\xi_b \tau_b P(0|0,\xi_b)} = A + B \tag{15}$$

Using this splitting of the integral, we can use the respective valid asymptotics $s\tau_b \ll 1, \xi_b \sim 1 - s\tau_b$ for $A$, and $s\tau_b \gg 1, \xi_b \sim \frac{1}{s\tau_b}$ for $B$ to obtain, as $s \to 0$ with $as \sim 1$

$$A \sim \int_0^{\tau^{-1}(1/s)} \frac{db}{\tau_b P(0|0, 1 - s\tau_b)}, B \sim s \int_{\tau^{-1}(1/s)}^a \frac{db}{P(0|0, \frac{1}{s\tau_b})} \sim as \tag{16}$$

so that

$$\hat{P}(0, a, s) \underset{s \to 0, as \sim 1}{\sim} e^{-as - \int_0^{\tau^{-1}(1/s)} \frac{db}{\tau_b P(0|0, 1 - s\tau_b)}} \tag{17}$$

In real space, the term $e^{-as}$ only shifts the time $t \mapsto t - a$. Therefore, the statistics of $P(0, a, t)$ for $a, t \to \infty, a \sim t$ only depend on the behavior of the integral $\int_0^{\tau^{-1}(1/s)} \frac{db}{\tau_b P(0|0, 1 - s\tau_b)}$.

## B. 1D LARW

Here we consider a $1D$, ergodicity breaking, non-persistent LARW. The study for a persistent LARW is similar. Using (17) with $P(0|0, 1 - s\tau_b)^{-1} \sim \sqrt{2s\tau_b}$, $\tau_b \gg b$, one obtains

$$\hat{P}(0, a, s) \underset{s \to 0, as \sim 1}{\sim} e^{-as - \sqrt{2s} \int_0^{\tau^{-1}(1/s)} \frac{db}{\sqrt{\tau_b}}} \tag{18}$$

Two cases arise : either $\int_0^\infty \frac{db}{\sqrt{\tau_b}} < \infty$, or this integral diverges. We begin with the former. In this case, we can write

$$\hat{P}(0, a, s) \underset{s \to 0, as \sim 1}{\sim} e^{-as - \sqrt{2s} \int_0^\infty \frac{db}{\sqrt{\tau_b}}} \tag{19}$$

so that, performing a Laplace inversion

$$P(0, a, t) \underset{a \sim t, t \to \infty}{\sim} \frac{I e^{-\frac{I^2}{2(t-a)}}}{\sqrt{2\pi}(t-a)^{3/2}}, I = \int_0^\infty \frac{db}{\sqrt{\tau_b}} \tag{20}$$

and writing $a = \gamma t, \gamma < 1$

$$P(0, a = \gamma t, t) \underset{t \to \infty}{\sim} t \times \frac{I e^{-\frac{I^2}{2(1-\gamma)t}}}{\sqrt{2\pi}\sqrt{(1-\gamma)^3 t^3}} = \frac{I e^{-\frac{I^2}{2(1-\gamma)t}}}{\sqrt{2\pi}\sqrt{(1-\gamma)^3 t}}, I = \int_0^\infty \frac{db}{\sqrt{\tau_b}} \tag{21}$$

This yields the value of the weight of surviving trajectories $w_d(t)$ defined in the main text

$$w_d(t) = 1 - \int_0^1 d\gamma \frac{I e^{-\frac{I^2}{2(1-\gamma)t}}}{\sqrt{2\pi}\sqrt{(1-\gamma)^3 t}} \tag{22}$$

If, however, $\int_0^\infty \frac{db}{\sqrt{\tau_b}} = \infty$, explicit Laplace inversion does not seem accessible.

### C. 2D LARW

We now consider a 2D LARW in the regime of ergodicity breaking. Using (17) and results from the main text, we obtain

$$\hat{P}(0, a, s) \underset{as \sim 1, s \to 0}{\sim} e^{-as - \pi \int_0^{\tau^{-1}(1/s)} \frac{db}{\tau_b \log \frac{8}{s\tau_b}}} \tag{23}$$

It can be shown that for $I = \int_0^\infty \frac{db}{\tau_b} < \infty$, for small $s$ one has the asymptotic behavior

$$\int_0^{\tau^{-1}(1/s)} \frac{db}{\tau_b \log \frac{8}{s\tau_b}} \underset{s \to 0}{\sim} \frac{I}{\log \frac{8}{s}} \tag{24}$$

Therefore, the Tauberian theorem along with the fact that the log varies very slowly leads to the explicit expression for the cumulative distribution of $\frac{a(t)}{t}$

$$\hat{P}_{cum}(a, s) = \frac{\hat{P}(0, a, s)}{s} \implies P\left(\frac{a(t)}{t} \geq \gamma\right) \underset{t \to \infty}{\sim} e^{-\frac{\pi I}{\log 8(1-\gamma)t}} \tag{25}$$

Therefore, the weight $w_d(t)$ for the 2D LARW is

$$w_d(t) \underset{t \to \infty}{\sim} 1 - P\left(\frac{a(t)}{t} \geq 0^+\right) \underset{t \to \infty}{\sim} 1 - e^{-\frac{\pi I}{\log 8t}}$$

### D. 3D LARW

In this section we consider a 3D ergodicity breaking LARW. Using again (17), one obtains, since $\int_0^\infty \frac{db}{\tau_b} < \infty$ by definition of the ergodicity breaking regime

$$\hat{P}(0, a, s) \underset{as \sim 1, s \to 0}{\sim} e^{-as - (1-R)I}, I = \int_0^\infty \frac{db}{\tau_b} \tag{26}$$

Therefore, we have for all $a$

$$P(0, a, t) \underset{t \to \infty}{\sim} \delta(t - a) e^{-(1-R)I} \tag{27}$$

In contrast to recurrent walks, the weight of the Dirac peak does not go to 1 and remainq stationary at long times.

## VIII. DISTRIBUTION OF THE ACTIVATION TIME FOR LARW

### A. The distribution of $a(t)$ as a marginal of the master equation

We consider as above a generic LARW on a lattice with a single activation site at the origin. A master equation can be written for the joint distribution of position and activation time, $P(x, a, t)$, taking into account the change of $\tau(a)$

$$\partial_t P(x, a, t) = \frac{1}{\tau(a)} \Delta_x P(x, a, t) - \delta_x \partial_a P(x, a, t)$$

where $\Delta$ is the discrete laplacian on the lattice and $\delta$ is the Kronecker delta. We did not use this equation in our derivation of the joint law of $x, a$ above, as we found it a less appropriate starting point for obtaining our results. However, it is well-suited for obtaining directly the marginal distribution of $a$. Summing the above over all sites of the lattice, one obtains the evolution equation for the distribution of $a$ written in the main text

$$\partial_t P(a, t) = -\partial_a P(0, a, t)$$

### B. Computing the cdf of $a$

In this section, we derive eq (18) in the main text. We focus on the non-ergodicity breaking case, which is $\tau(a) \ll a$, but we recall, as underlined in the main text, that this derivation is valid for trapped LARW with an *explicit* trapping threshold, i.e. with a diverging waiting time at a finite value $\tau(a_f) = \infty$. As proved above, the cdf of $a$ writes, after Laplace transformation $t \to s$,

$$\mathcal{L}_{t \to s} \mathbb{P}(a(t) \geq a) = \frac{\hat{P}(0, a, s)}{s} = \frac{\exp\left(-\int_0^a \frac{db}{\tau(b)\xi_b P(0|0,\xi_b)}\right)}{s} \tag{28}$$

The last equality follows from eq. (8) in the main text. Using $\tau(a) \ll a$, an asymptotic form of $P(0|0,\xi_a)$ for $as \ll 1$ can be derived and writes, depending on the space dimension

$$P(0|0,\xi_a) = \begin{cases} \frac{\xi_a}{\sqrt{1-\xi_a^2}} \underset{as \ll 1}{\sim} \frac{1}{\sqrt{2s\tau(a)}}, d = 1 \\ \frac{2}{\pi K(\xi_a^2)} \underset{as \ll 1}{\sim} \frac{1}{\pi} \log \frac{8}{s\tau(a)}, d = 2 \\ \underset{as \ll 1}{\to} \frac{1}{1-R}, d = 3 \end{cases} \tag{29}$$

Due to the algebraic $\frac{1}{s}$ factor and the slowly-varying functions appearing as arguments in the exponential term that we wrote above, eq. (28) is particularly well-suited for the use of a Tauberian theorem to obtain the long-time asymptotics of the cdf of $a$. The application of this theorem yields the cdf of $a$ upon Laplace inversion

$$\mathbb{P}(a(t) \geq a) \underset{t \to \infty, a \ll t}{\sim} \begin{cases} \mathrm{erfc}\left(\frac{\int_0^a \frac{db}{\sqrt{\tau(b)}}}{\sqrt{2t}}\right), d = 1 \\ \exp\left(-\frac{\pi \int_0^a \frac{db}{\tau(b)}}{\log 8t}\right), d = 2 \\ \exp\left(-(1-R)\int_0^a \frac{db}{\tau(b)}\right), d = 3 \end{cases} \tag{30}$$

## IX. NUMERICAL SIMULATIONS

To simulate LARWs, we use an event-driven Monte-Carlo algorithm. Each step in the algorithm represents a jump of the walker. To keep track of the time spent since the beginning of the walk, we have a simple rule :

- If the walker with activation time $a$ is on a site that is not an activation site, it jumps after a random time drawn from the exponential distribution with mean $\tau(a)$

- If the walker is on the activation site and arrived with activation $a_0$, then it jumps after a time $T$ that is drawn from the distribution with cumulative distribution $P_s(t)$. This cumulative distribution verifies

$$\frac{dP_s}{dt} = -\tau(a_0 + t)^{-1}$$

To draw $T$, we draw $y \in [0, 1]$ uniformly, so that $T = P_s^{-1}(y)$.

---



\end{document}